# Measurement of the transverse spatial quantum state of light at the single-photon level

**Brian J. Smith, Bryan Killett, M. G. Raymer**
*Oregon Center for Optics and Department of Physics, University of Oregon, Eugene, Oregon 97403*
**I. A. Walmsley**
*Clarendon Laboratory, University of Oxford, Parks Road, Oxford OX1 3PU, UK*
**K. Banaszek**
*Institute of Physics, Nicolaus Copernicus University, Grudziadzka 5, PL-87-100 Torun, Poland*

**Abstract:** We present an experimental method to measure the transverse spatial quantum state of an optical field in coordinate space at the single-photon level. The continuous-variable measurements are made with a photon-counting, parity-inverting Sagnac interferometer based on all-reflecting optics. The technique provides a large numerical aperture without distorting the shape of the wave front, does not introduce astigmatism, and allows for characterization of fully or partially coherent optical fields at the single-photon level. Measurements of the transverse spatial Wigner functions for highly attenuated coherent beams are presented and compared to theoretical predictions.
©2005 Optical Society of America
**OCIS codes:** (030.1640) Coherence; (110.1650) Coherence Imaging; (270.1670) Coherent Optical Effects

Individual light quanta, or photons, have a wealth of structure that has yet to be fully explored and utilized. Frequently, photons are used to carry information. Indeed, single photons are the basis of communications systems whose security is guaranteed by quantum mechanics. The most common coding is as "qubits", in which logical "0" and "1" are represented by two orthogonal polarization states, which form a discrete degree of freedom of the photonic quantum state. Currently, spatial degrees of freedom of single photons are being studied as a means to encode information [1,2], and thus a full characterization of arbitrary, continuous spatial states of photons is important. Furthermore, measurement of the spatial state of single photons puts the concept of the photon wave function [3,4] on solid ground. The state of photon polarization, which is a discrete degree of freedom, has been treated previously [5].

In this Letter we demonstrate a technique, shown in Fig.1, for measuring the continuous-variable, transverse spatial state of an optical field at the single-photon level. This technique is based on a method to directly measure the transverse spatial Wigner function (WF) of a light field using a common-path (Sagnac) interferometer and an area-integrating detector [6]. The measurement of the transverse spatial Wigner function for an ensemble of identically prepared photons completely characterizes the transverse spatial state of this ensemble. The Wigner function for a single-photon source is equivalent to the density matrix describing the quantum state, for either pure or mixed states. For pure states, the complex electric field, $E(\mathbf{x})$, up to a normalization constant, equals the photon wave function in coordinate space [3,4].

A key feature of the method is that every run generates a datum: it is not required to filter the input light, and thus discard photons. This requires the use of a detector that itself does not spatially or angularly filter the light, and is sensitive to single photons. In Ref[6] (hereafter referred to as MBWD), the design of the interferometer restricted the range of transverse wavevector and spatial extent of the input photonic quantum state, because the interferometer optics and proposed small-area detector, an avalanche photodiode (APD), limited the field stop and entrance pupil sizes.

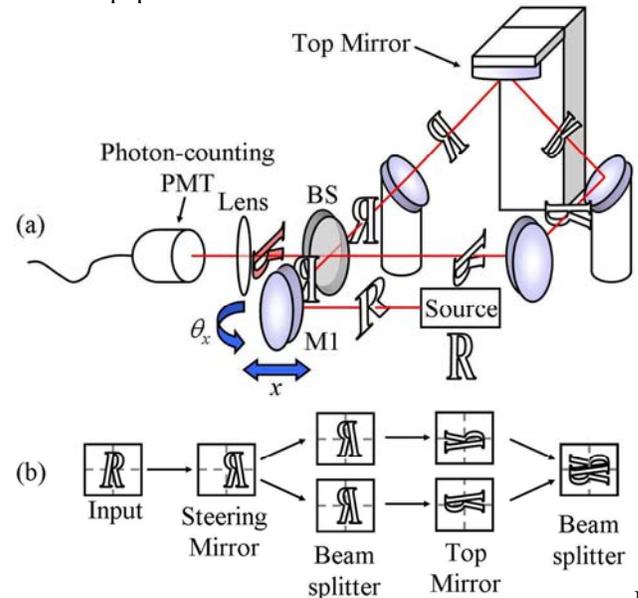

Fig. 1. (a) Experimental apparatus showing the top-mirror and wave-front rotation. Only the wave-front transformations along the clock-wise path are depicted. The external steering mirror M1 tilts $(\theta_x)$ and translates $(x)$. (b) Wave front transformations after optical elements of the interferometer.

The method that we use to measure the photon's continuous-variable state is a generalization of that proposed in MBWD, with one crucial difference and other substantive improvements. In the MBWD scheme, a light beam is sent into a Sagnac interferometer where it is split at a beam splitter. A Dove prism in the interferometer performs a relative rotation of 180° and a mirror inversion on the wave fronts of the two counter-propagating beams. In effect, the Dove prism performs a two-dimensional parity operation on one of the beams relative to the other. The fields are recombined at the beam splitter and interfered on a large-area p-i-n photodiode. The mean photocurrent is directly proportional to the transverse spatial Wigner function at a point in phase-space that is set by the tilt and translation of a mirror external to the interferometer. Use of a Dove prism inherently introduces astigmatism to the optical beams and can lead to measurement errors. Furthermore, the ability of the interferometer to collect divergent light, which is measured by the device's numerical aperture (NA $\approx$ 0.01), is severely limited by the cross section of the Dove prism. A low numerical aperture not only limits the amount of light collected, but also lowers the diffraction-limited spatial resolution. The need for a large-area detector arises inherently from the intrinsic divergence of the signal field and the need to scan the tilt and location of the external mirror. This breaks the beam emerging from the Sagnac into two distinct beams that cannot be imaged to a single small-area detector such as an APD. These conditions make the MBWD scheme best suited to characterizing nearly collimated light beams with large intensities.

The improved scheme shown in Fig.1. employs an all-reflecting, parity-inverting Sagnac interferometer, and a large-area photon-counting photomultiplier tube (PMT), allowing the emerging beams to be detected at the single-photon level. By eliminating the Dove prism, we remove astigmatism and increase the numerical aperture of the device to an observed value of 0.09, and potentially to as high as 0.3. Light from the source is directed by an external steering mirror, M1, into the interferometer. The field is split by a 50:50 beam splitter into clockwise and counter-clockwise propagating beams. As the two beams pass through the interferometer, they are directed out of the plane of the table by three mirrors, in what we call a "top-mirror configuration." The angles between all beam propagation directions in the interferometer are 90°.

This arrangement rotates the wave fronts of the two beams by ±90° for the counter-clockwise and clockwise directions, respectively, and inverts them along the horizontal giving the transformations $E(x,y) \rightarrow E(\pm y, \pm x)$. The opposition of the rotations is due to the direction of propagation through the apparatus, and can be understood using a Berry's phase argument [7]. This effectively performs a two-dimensional parity operation on one of the wave fronts while leaving the other unchanged. (The optical polarizations are modified only slightly.) The beams are then recombined at the beam splitter and collected with a lens onto a large-area, photon-counting photomultiplier tube. The average count rate from the PMT is proportional to a constant term plus a term proportional to the Wigner function [6].

That the parity-inverting Sagnac interferometer measures the WF can be seen by expressing the WF of a state $\hat{\rho}$ as the expectation value of the parity operator $\hat{\Pi}$, evaluated with a state that has been displaced in phase space by a transverse distance $\mathbf{x}$ and transverse wave vector $\mathbf{k}$, [8]

$$W(\mathbf{x},\mathbf{k}) = \frac{1}{\pi^2} Tr\left[\hat{D}^{-1}(\mathbf{x},\mathbf{k})\hat{\rho}\hat{D}(\mathbf{x},\mathbf{k})\hat{\Pi}\right] \quad (1)$$

where $\hat{D}(\mathbf{x},\mathbf{k})$ is the phase-space displacement operator. For a complex quasi-monochromatic scalar field, $E(\mathbf{x})$, the transverse spatial Wigner function equals

$$W(\mathbf{x},\mathbf{k}) = \frac{1}{\pi^2} \int d^2 x' \langle E(\mathbf{x}+\mathbf{x}')E^*(\mathbf{x}-\mathbf{x}')\rangle e^{2i\mathbf{k}\cdot\mathbf{x}'} \quad (2)$$

where $\mathbf{x}' = (x', y')$ is the two-dimensional vector in the object plane, $\mathbf{k} = (k_x, k_y)$ is the transverse-spatial wave vector, and the angle brackets imply an ensemble average over all statistical realizations of the field. For a single-photon field, $\langle E(\mathbf{x}_1)E^*(\mathbf{x}_2)\rangle$ is the density matrix and represents the quantum state of the photon.

As pointed out by MBWD, the average count rate from the photomultiplier tube is proportional to three terms: $\langle N \rangle = \langle n_1 \rangle + \langle n_2 \rangle + \langle n_{12} \rangle$. The first two terms are proportional to the intensities of the individual counter-propagating beams, which are independent of the position and tilt of the steering mirror, M1. The final term is proportional to the transverse spatial WF, with arguments determined by the position $\mathbf{x} = (x, y)$ and tilt angles $\theta_i = \arcsin(k_i / k_0)$, (subscript $i = x$ or $y$ and $k_0$ is the wavenumber), of the steering mirror, M1,

$$\langle n_{12} \rangle = -\eta(\pi^2/2)W(\mathbf{x},\mathbf{k}) \quad (3)$$

Here $\eta$ is related to the detector's efficiency and the negative is due to the reflection phase shift from the beam splitter.

A Fresnel diffraction calculation shows that the measured WF describes the spatial state of the field at the plane in which the beams are tilted, that is, in the plane of the steering mirror, M1. This observation corrects a statement in MBWD.

The interferometer was constructed with 51mm-diameter dielectric mirrors and a 51mm-diameter non-polarizing beam splitter. The steering mirror, M1, was mounted on translation and rotation stages. The light at the output of the interferometer was collected using a 51mm-diameter lens, of 75mm focal length, onto the large-area, photon-counting PMT module (Hamamatsu H7422-50) with a quantum efficiency of approximately 11% at 633nm, and spatial uniformity > 85%. A computer controlled the steering mirror position via two actuators and collected the average photon count rate at each position.

To demonstrate the ability of this interferometer to measure transverse spatial Wigner functions at the single-photon level, we examined several fields derived from a highly attenuated He-Ne laser. For simplicity, only one-dimensional fields are considered, that is, fields that vary in only one transverse spatial dimension, $x$. This can be accomplished by using fields that can be factored into functions that depend on orthogonal coordinates $(x, y)$.

Measured transverse spatial Wigner functions are shown in Fig.2. These were obtained using fields attenuated such that there was less than a single photon present in the interferometer on average. The photon-counting rate was typically centered about 100,000counts/s. The first plot, (a), depicts the transverse spatial WF for a "top-hat" field distribution. This field was created by directing an expanded laser beam onto a single slit (width = 0.40mm) placed just before the steering mirror. The triangular peak along the $k_x = 0$ section represents the autocorrelation of the top-hat function, and the large values of $k_x$ near the edges correspond to diffraction from the edges of the slit. The second plot (b) depicts the transverse spatial WF of two top-hat fields displaced transversely from one another. This field was created by passing an expanded laser beam through a double slit (spacing = 0.28mm and width = 0.06mm) placed just before the steering mirror. The plot shows two lobes displaced along the $x$-axis by an amount equal to the slit spacing (0.28mm) and an oscillating region in between, which is a signature of the mutual coherence of the two top-hat fields. The measured WFs are in excellent qualitative agreement with theoretical models. The measured slit width of the top-hat field matches the actual slit width (0.40mm) and the divergence angle given by $k_{x-\max}/k_0$ matches the predicted value 1.51mrad. Similarly the measurement of the slit widths (0.06mm), double-slit spacing (0.28mm), and divergence angles for each slit (6.0mrad) match the actual data as well. The plots are constructed from the raw data by subtracting off the constant terms, multiplying by negative one, and numerically normalizing.

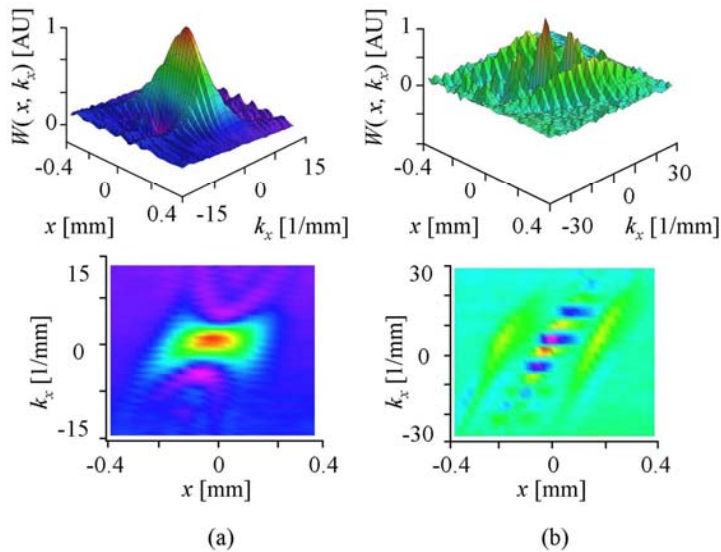

Fig. 2. Single-photon level transverse spatial Wigner function measurements for (a) the "top-hat" field distribution and (b) the double slit field distribution.

There are several areas of current research that would benefit from measurements of the spatial state of non-collimated optical fields at the single-photon level. The field of quantum information processing will most likely look to photons to carry information from processor to processor. These schemes rely on the subtle interference of non-classical states of light. This will require near perfect overlap of their spatial states, making characterization essential. Similarly, free-space quantum communication, including key distribution, relies on the preservation of qubit properties during propagation of single photons.

Beyond characterization of single-photon fields, spatially entangled pairs of photons, such as those produced through spontaneous parametric down conversion, contain a wealth of quantum information. Such a system exhibits continuous-variable entanglement (position and momentum), analogous to the quadratures of a single-mode field. With such quantum objects one can realize a form of the original EPR type of continuous-variable entanglement. Such entangled photon pairs are the basis of several novel experiments such as tests of quantum nonlocality [9], quantum imaging [10], and quantum lithography [11].

By sending two spatially-entangled photons into two parity-inverting interferometers, one can measure the joint two-photon transverse spatial Wigner function, and completely characterize the transverse entanglement of this system [6]. With this technique, entanglement can be fully characterized. Furthermore, the study of spatial decoherence of classical waves passing through a random scattering medium [12] can then be extended to the quantum realm by observing photon disentanglement. Understanding such disentanglement is not only of fundamental interest, but plays a key role in the efficient application of photons as quantum information carriers.

In conclusion, we present the first measurements of the continuous-variable transverse spatial state of light at the single-photon level. The scheme developed is particularly geared toward measurement of single photons, including those scattered in random media. This technique adds to and improves upon the previous methods of optical field measurement. The phase-space tomography [13] and two-point correlation function approaches [12, 14, 15], for example, were limited to fields of at least a few photons. The new technique should be of use for characterizing extremely low-intensity light fields, and serve as a probe for spatial decoherence and entanglement in optical systems.

The authors thank Andrew Nahlik for helpful contributions. This research was supported by the National Science Foundation's ITR and INT programs, grant nos. 0219460 and 0334590. B. J. Smith's e-mail address is bsmith4@uoregon.edu.